\documentclass[twocolumn]{aastex63}
\usepackage{natbib}
\usepackage{multirow}
\newcommand\mname{DYNAMITE}
\newcommand{\tauc}{\object{$\tau$\,Ceti}}

\received{}
\revised{}
\accepted{}
\submitjournal{AJ}

\shorttitle{\tauc{} System}
\shortauthors{Dietrich \& Apai}
\graphicspath{{./}{}}

\begin{document}

\defcitealias{tuo13}{T13}
\defcitealias{fen17}{F17}

\title{An Integrated Analysis with Predictions on the Architecture of the \tauc{} Planetary System, Including a Habitable Zone Planet}

\correspondingauthor{Jeremy Dietrich}
\email{jdietrich1@email.arizona.edu}

\author[0000-0001-6320-7410]{Jeremy Dietrich}
\affiliation{Department of Astronomy, The University of Arizona, Tucson, AZ 85721, USA}

\author[0000-0003-3714-5855]{D\'aniel Apai}
\affiliation{Department of Astronomy, The University of Arizona, Tucson, AZ 85721, USA}
\affiliation{Lunar and Planetary Laboratory, The University of Arizona, Tucson, AZ 85721, USA}



\begin{abstract}

\tauc{} is the closest single Sun-like star to the solar system and hosts a multi-planet system with four confirmed planets.  The possible presence of additional planets, especially potentially habitable worlds, remains of great interest.  We analyze the structure of the \tauc{} planetary system via the \mname{} algorithm, combining information from exoplanet population statistics and orbital dynamics with measurements of this specific system.  We also expand \mname{} to incorporate radial velocity information.  Our analysis suggests the presence of four additional planets, three of which match closely with the periods of three tentative planet candidates reported previously.  We also predict at least one more planet candidate with an orbital period between $\sim270-470$ days, in the habitable zone for \tauc{}.  Based on the measured $m \sin i$ values of the confirmed planets, we also assess the possible masses and nature of the detected and undetected planets.  The least massive planets and candidates are likely to be rocky, while the other planets and candidates could either be rocky or contain a significant gaseous envelope.  The RV observable signature from the predicted habitable zone planet candidate would likely be at or just above the noise level in current data, but should be detectable in future extremely high-precision radial velocity and direct imaging studies.

\end{abstract}



\section{Introduction} \label{sec:intro}

As a lone G8V dwarf located $3.650\pm0.002$ pc from the Sun \citep[e.g.,][]{tei09}, \tauc{} has been extensively studied since the early 1900s.  It is the second-closest star similar to our Sun after $\alpha$ Centauri A, and the closest single Sun-like star \citep[e.g.,][]{hal04}.  It was noted as early as 1916 that \tauc{} was a G dwarf with parallax of 320 mas \citep{ada16}, an overestimate of only 17\% from the current measured value.  In the 1950s \tauc{} was theorized to be one of only two other stars within 5~pc of the Sun capable of supporting life in a hypothetical planetary system \citep{hua59}, and therefore was targeted in Frank Drake's Project Ozma with radio telescopes to search for repeated signals from advanced civilizations \citep{dra61}.  As planet hunting became more and more sophisticated, \tauc{} remained one of the favorite targets due to its similarity to our Sun.

Beyond (or possibly due to) its astronomical significance, \tauc{} is one of the few stars that plays important roles in popular culture.  It has been prominently featured in science fiction novels as a home to extraterrestrial civilizations by the likes of Isaac Asimov, Robert Heinlein, Ursula K.  Le Guin, and Arthur C.  Clarke.  Multiple episodes of science-fiction TV show standards \textit{Star Trek} and \textit{Doctor Who} have referenced a planetary system with habitable worlds orbiting \tauc{}.  All of these were created before any planets were discovered in the \tauc{} system.

Given its importance within and beyond astronomy, the planetary architecture of the \tauc{} system and its prospect for hosting a habitable world is of broad interest.  As of yet, however, this nearby planetary system has only been partially explored.  Thus, the question emerges: What other worlds may be present in \tauc{}, where would these be located within the system, and would these planets be detectable?

The \mname{}\footnote{\url{https://github.com/JeremyDietrich/dynamite}} \citep{die20} algorithm can predict the presence and parameters of currently unknown planets in multi-planet systems.  \mname{} combines specific -- but often uncertain and incomplete -- information on the given planetary system with robust, exoplanet population-level information and first-principles-based orbital dynamical considerations, an approach that can provide enhanced statistical understanding of individual planetary systems \citep[see also][]{bix17,apa18}.  This integrated approach was recently applied to over forty TESS-discovered multi-planet systems \citep{die20}.  We will explore the \tauc{} system to assess contextual evidence for the presence of additional planets and guide searches for them.  Furthermore, as the known planets e and f straddle the optimistic limits for the habitable zone around \tauc{} \citep[e.g.,][]{kop13}, an integrative study can help statistical assessment of the presence of a potential habitable planet around the closest single Sun-like star.

This paper is organized as follows.  Section~\ref{sec:tauc} details the \tauc{} planetary system as well as its stellar properties, and we briefly introduce the assumptions in \mname{} used specifically for \tauc{} in Section~\ref{sec:methods}.  Section~\ref{sec:results} presents the \mname{} predictions for the system.  Finally, Section~\ref{sec:discussion} discusses the effect of our assumptions on the results and the supporting statistical evidence for planet candidates reported, as well as exploring the nature of the planets (including their possible habitability) and assessing the predicted observational signatures for unknown candidates that can be tested in the near future.

\section{The \texorpdfstring{$\tau$\,}{tau }Ceti System}\label{sec:tauc}

\begin{figure*}[ht]
    \centering
    \includegraphics[width=2.12\columnwidth]{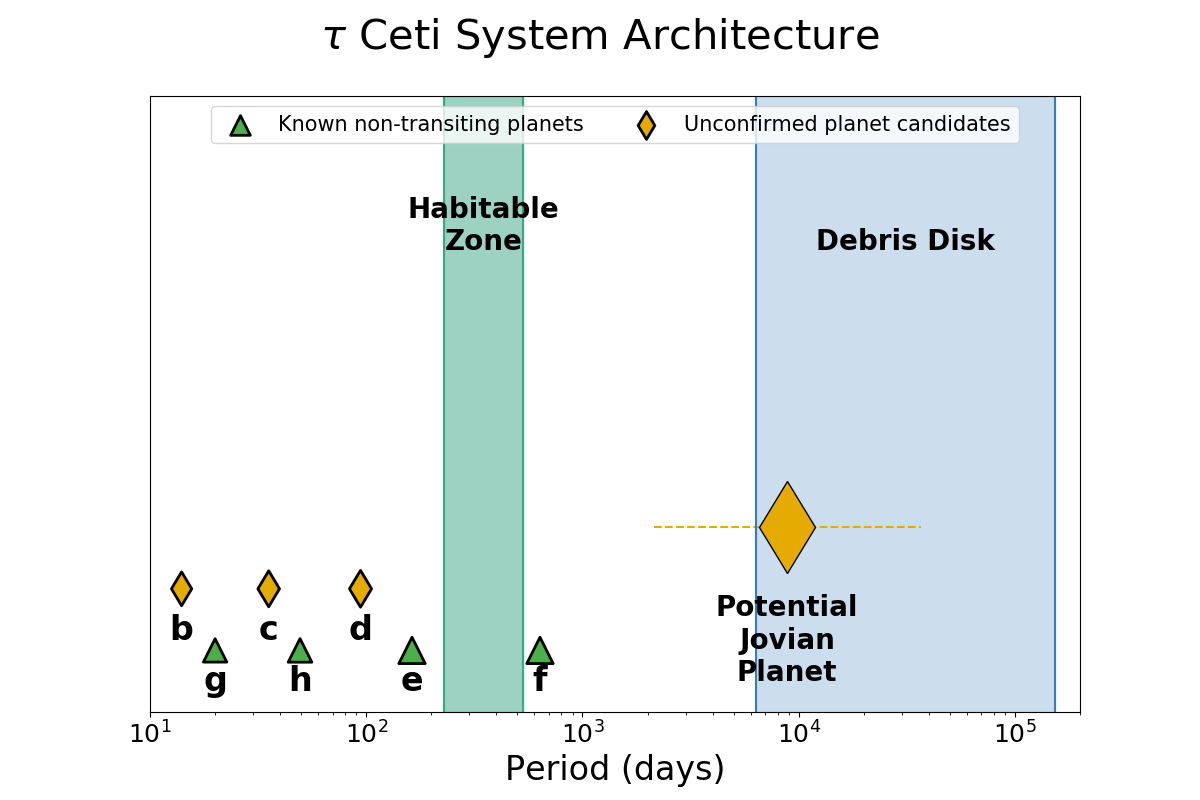}
    \caption{The \tauc{} system architecture, with known planets, unconfirmed planet candidates (with large error bars for the tentative Jovian planet detection), and the extents of the habitable zone and the debris disk.  Relative marker sizes match planet sizes.}
    \label{fig:tCa}
\end{figure*}

\tauc{} is a G-spectral type, main-sequence star that is somewhat smaller than the Sun and slightly less than half as luminous.  It is also less active than the Sun, with not quite as strong a starspot cycle; observations of the rotation and activity cycle of \tauc{} argue for a nearly pole-on (i.e., low-inclination) configuration \citep[e.g.,][]{gra94}.  Relevant stellar parameters for \tauc{} can be found in Table~\ref{tab:tauc}, and the known system architecture (as stated in the following paragraphs) is shown in Figure~\ref{fig:tCa}.

\begin{table}[ht]
    {\centering
    \caption{Stellar parameters for \tauc{}}
    \label{tab:tauc}
    \begin{tabular}{|l|c|r|}
        \hline
        Parameter Name & Value & Ref.\\
        \hline
        Spectral Type & G8V & (a)\\
        Mass ($M_\odot$) & $0.783\pm0.003$ & (b)\\
        Radius ($R_\odot$) & $0.793\pm0.004$ & (b)\\
        Luminosity ($L_\odot$) & $0.448\pm0.010$ & (b, c)\\
        Temperature (K) & $5344\pm50$ & (c)\\
        Distance (pc) & $3.650\pm0.002$ & (b)\\
        Rotation period (d) & 34 & (d)\\
        Age (Gyr) & 5.8 & (a)\\
        \hline
    \end{tabular}
    }
    \\[10pt]
    \textbf{Notes}: (a) \citet{mam08}, (b) \citet{tei09}, (c) \citet{san04}, (d) \citet{bal96}.  The rotation period given by \citet{bal96} and the age given by \citet{mam08} have no attached uncertainties.  We do not report Gaia DR2 values for distance and radius, as those have higher uncertainties due to pixel saturation \citep{ker19}.
\end{table}

Radial velocity (RV) data of \tauc{} collected over a decade contain multiple periodic signals with periods ranging from 14 to 630 days.  The analyses of the periodic RV modulations argue for 3 unconfirmed planet candidates with periods of 14 (b), 35.4 (c), and 94.1 (d) days \citep[][hereafter \citetalias{tuo13}]{tuo13} and 4 planets with periods of 20 (g), 49.4 (h), 163 (e), and 636 (f) days \citep[\citetalias{tuo13};][hereafter \citetalias{fen17}]{fen17}.  In particular, \citetalias{fen17} confirmed planets e and f and detailed two strong signals they attribute to  planets g and h.  However, while \citetalias{fen17} did find marginal evidence for each of the planet candidates labeled b, c, and d in the study by \citetalias{tuo13}, the signals were not strong enough to confirm these planets (see Table~\ref{tab:planets} for a summary of the known planet properties).

In addition to the RV-detected planets and planet candidates, there is also tentative evidence from astrometric measurements for the presence of an additional, long-period, very massive planet.  A tangential velocity anomaly in the HIPPARCOS astrometric proper motion measurements of \tauc{} can be interpreted as a Jovian-mass planet in a further-out orbit ($1-2 M_{Jup}$ at 3-20~au; \citealt{ker19}).  We do not include this candidate in our study, as the constraints on the planet candidate are too loose to meaningfully inform our analysis.  However, we note that the inner edge of its semi-major axis range would be near a 4:1 period ratio with planet f, and would therefore provide an interesting region of period space to explore further.

\tauc{} is also known to host a debris disk, and its inclination was inferred to be 35$\pm10^\circ$ from the best-fit model to observations taken using the \textit{Herschel Space Observatory} and the Atacama Large Millimeter Array (ALMA) \citep{law14, mac16}.  The comparison of the disk's inferred inclination and the star's inferred inclination suggest that the disk plane is in or close to the star's equatorial plane.  This also implies that the planets in the system may, too, orbit at low-inclination orbits, seen nearly face-on from our vantage point.

\section{Methods} \label{sec:methods}

In this study we use \mname{} to integrate the observational evidence on \tauc{} (considering its uncertainties and incompleteness) with statistical, population-level constraints on exoplanet populations (orbital period distributions, planetary architectures, planet size/mass distributions) and with orbital dynamical stability considerations.  For a full description of \mname{} we refer to \citet{die20}, and we only briefly review here the principles underpinning \mname{}.

The goal of \mname{} is to integrate the available specific and statistical information to predict the likelihood distributions for orbital period, planet radius, and inclination for yet-undiscovered planets.  \mname{} utilizes a Monte Carlo implementation.  As detailed below, \mname{} draws from intrinsic planet size, orbital period, and inclination distributions, as established from the Kepler planet population statistics \citep[e.g.,][]{mul18}.  Putative planets drawn from these distributions are further analyzed for orbital stability: it has been shown \citep[e.g.,][]{hef19} that long-term stability of pairwise planets in a system requires 8-10 mutual Hill radii separation.  Therefore, we limit the orbital placements of planets to distances greater than 8 mutual Hill radii of each other.  This is only an approximate necessary condition, and while it works for current populations, future studies could utilize a more precise formalism based on a fuller dynamical stability assessment (i.e., including non-pairwise or three-planet mutual interactions) of each manifestation of the system studied.

In the current implementation of \mname{}, we use two prescriptions for the planets' orbital period probability distribution.  In this study we explore and contrast these two, currently equally valid prescriptions.  The first prescription comes from the analysis of the Kepler planet population via the Exoplanet Population Observation Simulator \citep[EPOS;][]{mul18}.  This study finds that planets are more likely to have similar period ratios between them, with each period ratio being drawn from a Log-normal distribution.  The second comes from the Exoplanets Systems Simulator \citep[SysSim;][]{hef19}.  This study finds that planets tend to be clustered in period space, with the probability of a specific period inside a cluster also determined by a Log-normal distribution.

The population statistics come from the Kepler dataset, and \mname{} was tested on both Kepler and TESS multi-planet systems \citep[for details, see][]{die20}, showing a high degree of accuracy in predicting the parameters.  For transiting exoplanet systems, like those found by Kepler and TESS, the inclinations can be constrained via the transit impact parameter.  In addition, planet radii are known from the transit depth and the stellar radius.  In cases of non-transiting planets, however, where the planet radius is not known (i.e., for RV-detected exoplanets), we use the planet mass as a fundamental parameter instead of planet radius.  For these RV-detected systems, the planet mass is often degenerate with the orbital inclination unless another type of observation can constrain the system's geometry.

The planets and planet candidates in the \tauc{} system have been identified via RV observations and, thus, in the following we will use planet mass as a fundamental parameter.  While the earlier study by \citetalias{tuo13} identified five planets (b, c, d, e, and f), the subsequent and more comprehensive follow-up study by \citetalias{fen17} did not unequivocally confirm three of these (b, c, d).  Therefore, in our study planets b, c, and d will be considered as planet {\em candidates}, and are not provided as priors to our \mname{} modeling.  We do, however, use the robustly detected four planets (g, h, e, and f; \citealt{fen17}) as input.  

We also assume that the planets are closely co-planar and that their orbital plane is aligned (within a few degrees) with the inclination of the debris disk itself.  This assumption is reasonable as such planetary orbit and debris disk plane alignments have been reported for a number of systems \citep[e.g.,][]{apa15,pla20}, while -- to our knowledge -- no system with significant misalignment between debris disk and planetary orbital planes have been found.  Under the assumption that the planetary orbits' inclinations match that of the debris disk, the planets' true masses would then be approximately 1.75 times the $m \sin i$ values measured via RV observations.  We explore the effect of the assumption of relative co-planarity between the planets and the disk on the results of the analysis by testing a normal distribution for the debris disk inclination centered on the value from the best-fit model.  We find that the above assumption only increases the width of the confidence interval for the predicted planet masses, but does not change the mean value.  We also find that there would be no significant difference in the predicted planetary architecture with mutual inclinations between the planets and the disk $\lesssim 17^\circ$.  If the planets had a lower inclination than the disk (more face-on), we find from our dynamical stability model that the closest planet and candidate pairs would become unstable at $17^\circ$ less than the disk value.  If the planets had a higher inclination than the disk (more edge-on), every planet and candidate would likely be rocky at inclinations $17^\circ$ greater than the disk value.  Therefore, we state that this assumption of relative co-planarity does not have a major impact on the predicted planetary architecture of the \tauc{} system.

We use the possible masses of the known and predicted planets to constrain their nature and radii, via a mass--radius (M--R) relationship and by comparing their predicted mass/radii to such regimes as identified in the exoplanet population.  As mass--radius relationships are not yet fully understood, in our study we compare results based on two mass--radius relationships: a non-parametric and a power-law-based.  Specifically, we use the non-parametric M--R relationship by \citet{nin18} and the power-law mass-radius relationship characterized by \citet{ote20}, using both the ``rocky" and ``volatile-rich" populations.  These two planet populations overlap between $5-25 M_\oplus$, so we test the predictions of both populations in that mass range separately.  Notably, the predictions of their power-law relationships fit their entire dataset within 2$\sigma$, and match well with ensemble results predicted from planet formation models \citep[e.g., the Generation III Bern models;][]{ems20a, ems20b}.

\section{Results} \label{sec:results}

We report the properties of the planets and candidates, along with our predictions, in Table~\ref{tab:planets}.  The planet radii are not currently observable parameters for this system, as none of the planets transit and are too faint and/or too close to the host star for current direct imaging techniques.  Therefore, we report planet $m \sin i$ values determined from the planet radius distribution.  We report results from our analysis based on different assumptions on the two different orbital period distributions and on three different mass--radius relationships.  We find that the results from the two orbital period distributions are very similar, and -- for most planets -- the derived planetary natures under the three different M--R relationships are similar, too.

\begin{table*}[ht]
    \centering
    \caption{\tauc{} Planet and Planet Candidate Parameters}
    \label{tab:planets}
    \begin{tabular}{|l|c|c|c|c|r|}
        \hline
        \textbf{Name} & \textbf{Period (days)} & $\mathbf{m \sin i}$ & \textbf{Planet Type} & \textbf{Note} & \textbf{Origin/Reference} \\
        \hline
        \multirow{6}{*}{PxP--1} & \multirow{3}{*}{$12.0\:[7.10,\,13.9]$} & \multirow{6}{*}{$2.0\:\pm\:0.8$} & sub-Neptune & \multirow{6}{*}{Equivalent to b?} & Period Ratio, NP\\
         & & & super-Earth & & Period Ratio, ``Rocky"\\
         & & & super-Earth & & Period Ratio, ``Volatile"\\
        \cline{2-2}
         & \multirow{3}{*}{$14.0\:[10.0,\,14.8]$} & & sub-Neptune & & Clustered Periods, NP\\
         & & & super-Earth & & Clustered Periods, ``Rocky"\\
         & & & super-Earth & & Clustered Periods, ``Volatile"\\
         \hline
        \tauc{} b & $14.0_{-0.024}^{+0.017}$ & $2.0\pm0.8$ & Likely rocky & Unconfirmed Candidate & \citetalias{tuo13}\\
        \hline
        \tauc{} g & $20.0_{-0.01}^{+0.02}$ & $1.75_{-0.40}^{+0.25}$ & Likely rocky & Planet & \citetalias{fen17}\\
        \hline
        \multirow{6}{*}{PxP--2} & \multirow{3}{*}{$31.4\:[28.8,\,35.1]$} & \multirow{6}{*}{$3.1\:\pm\:1.4$} & sub-Neptune & \multirow{6}{*}{Equivalent to c?} & Period Ratio, NP\\
         & & & super-Earth & & Period Ratio, ``Rocky"\\
         & & & sub-Neptune & & Period Ratio, ``Volatile"\\
         \cline{2-2}
         & \multirow{3}{*}{$34.0\:[28.0,\,36.4]$} & & sub-Neptune & & Clustered Periods, NP\\
         & & & super-Earth & & Clustered Periods, ``Rocky"\\
         & & & sub-Neptune & & Clustered Periods, ``Volatile"\\
         \hline
        \tauc{} c & $35.4_{-0.106}^{+0.088}$ & $3.1\pm1.4$ & Unknown & Unconfirmed Candidate & \citetalias{tuo13}\\
        \hline
        \tauc{} h & $49.4_{-0.10}^{+0.08}$ & $1.83_{-0.26}^{+0.68}$ & Likely rocky & Planet & \citetalias{fen17}\\
        \hline
        \multirow{6}{*}{PxP--3} & \multirow{3}{*}{$89.7\:[78.7,\,105]$} & \multirow{6}{*}{$3.6\:\pm\:1.7$} & sub-Neptune & \multirow{6}{*}{Equivalent to d?} & Period Ratio, NP\\
         & & & super-Earth & & Period Ratio, ``Rocky"\\
         & & & sub-Neptune & & Period Ratio, ``Volatile"\\
         \cline{2-2}
         & \multirow{3}{*}{$67.0\:[67.0, 103]$} & & sub-Neptune & & Clustered Periods, NP\\
         & & & super-Earth & & Clustered Periods, ``Rocky"\\
         & & & sub-Neptune & & Clustered Periods, ``Volatile"\\
        \hline
        \tauc{} d & $94.1\pm0.7$ & $3.6\pm1.7$ & Unknown & Unconfirmed Candidate & \citetalias{tuo13}\\
        \hline
        \tauc{} e & $163_{-0.460}^{+1.08}$ & $3.93_{-0.64}^{+0.83}$ & Unknown & Planet & \citetalias{tuo13, fen17}\\
        \hline
        \multirow{6}{*}{PxP--4} & \multirow{3}{*}{$322\:[277,\,395]$} & $3.6\:[2.70,\,3.97]$ & sub-Neptune & \multirow{6}{*}{No equivalence} & Period Ratio, NP\\
         & & $1.91\:[0.912,\,6.14]$ & super-Earth & & Period Ratio, ``Rocky"\\
         & & $1.91\:[0.941,\,3.37]$ & super-Earth & & Period Ratio, ``Volatile"\\
         \cline{2-2}
         & \multirow{3}{*}{$468\:[406,\,468]$} & $3.6\:[2.70,\,3.97]$ & sub-Neptune & & Clustered Periods, NP\\
         & & $1.91\:[0.912,\,6.14]$ & super-Earth & & Clustered Periods, ``Rocky"\\
         & & $1.91\:[0.941,\,3.37]$ & super-Earth & & Clustered Periods, ``Volatile"\\
        \hline
        \tauc{} f & $636_{-47.7}^{+11.7}$ & $3.93_{-1.37}^{+1.05}$ & Unknown & Planet & \citetalias{tuo13, fen17}\\
        \hline
    \end{tabular}
    \\[10pt]
    \textbf{Notes}: \textit{``Name"}: the given planet designations for both the confirmed planets and the candidates, as well as the planets we predict.  PxP stands for ``\textbf{P}redicted e\textbf{x}o\textbf{P}lanet". \textit{``Period"}: The orbital period and uncertainty for the known planets and candidates, compared to our predicted value and 16\%-84\% confidence interval from \mname{}. ``\textit{m} sin \textit{i}": the minimum mass of the planets, candidates, and predictions, not assuming an inclination.  For PxP--1-3, we give the \textit{m} sin \textit{i} as the same value of the equivalent known planet candidate. \textit{``Planet Type"}: The predicted planet type for the PxPs and the natures for the known planets, all assuming an orbital inclination of 35$^\circ$. \textit{``Note"}: Provides comparison between PxPs and known planets/candidates. \textit{``Origin/Reference"}: Provides the reference paper for the confirmed planets and candidates, and the specific population models in \mname{}.  NP is the non-parametric M--R relationship from \citet{nin18}, ``Rocky" and ``Volatile" are the rocky and volatile-rich populations in the \citet{ote20} power-law M--R relationship.  Each line in the table for each PxP corresponds to a different combination of population models, and repeated values are combined together.
\end{table*}

\subsection{Period Distribution for Equal Period Ratios}

Under the assumption that planets are found with equal period ratios between them, the \mname{} predictions agree very well with the combined picture from the analyses by \citetalias{tuo13} and \citetalias{fen17}.  Specifically, \mname{} predicts four planet candidates (PxP--1 through PxP--4, see Table~\ref{tab:planets}) with orbital periods at the relative likelihood maxima of 12.0, 31.4, 89.7, and 322 days.  We point out that the planet candidates b, c, and d reported by \citetalias{tuo13} (which were not part of the \mname{} input) have very similar periods (b = 14.0 days, c = 35.4 days, d = 94.1 days) to the predicted planets PxP--1, PxP--2, and PxP--3.  Combining the four planets from \citetalias{fen17} with the three unconfirmed (but now supported) candidates and the prediction of PxP--4 (which has no current equivalent candidate in the observational literature), the \tauc{} system becomes strongly dynamically packed.  The relative likelihood in log-period space for the period ratio prescription is shown in the top of Figure~\ref{fig:tCP}.

\subsection{Period Distribution for Clustered Periods}

\begin{figure*}[ht]
    \centering
    \includegraphics[width=1.92\columnwidth]{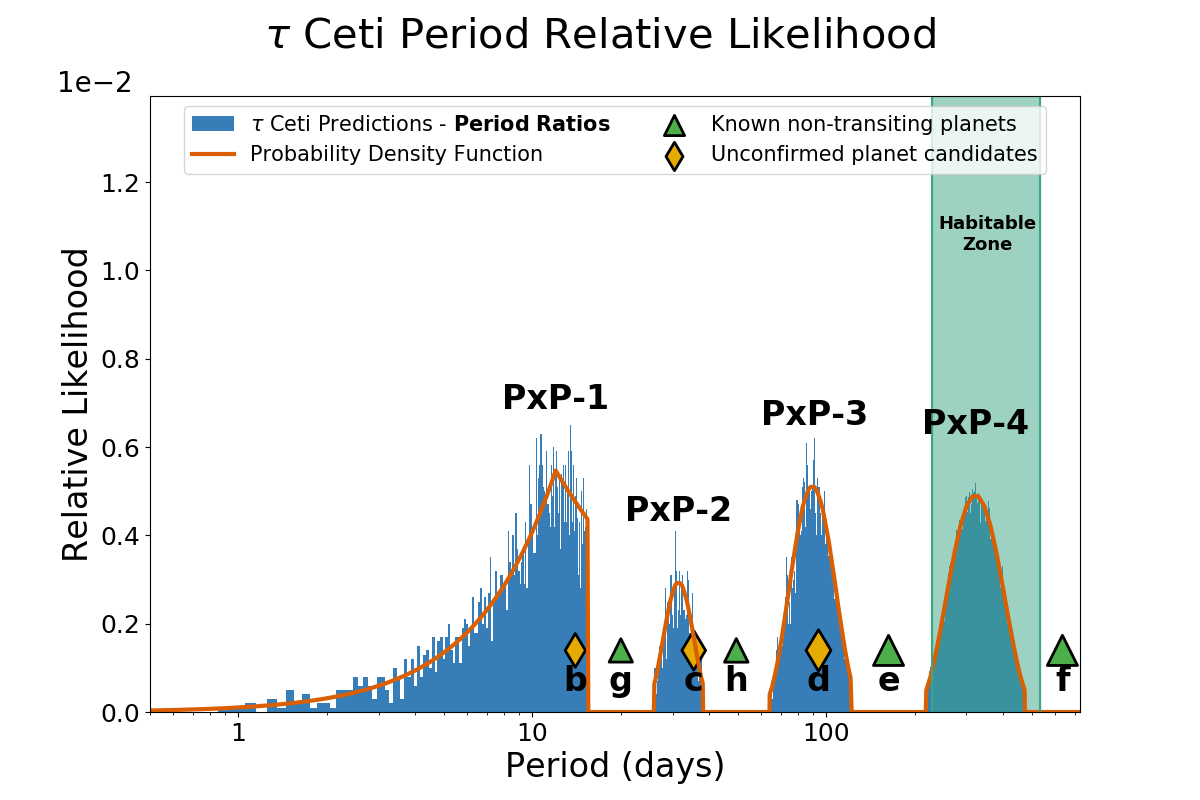}
    \includegraphics[width=1.92\columnwidth]{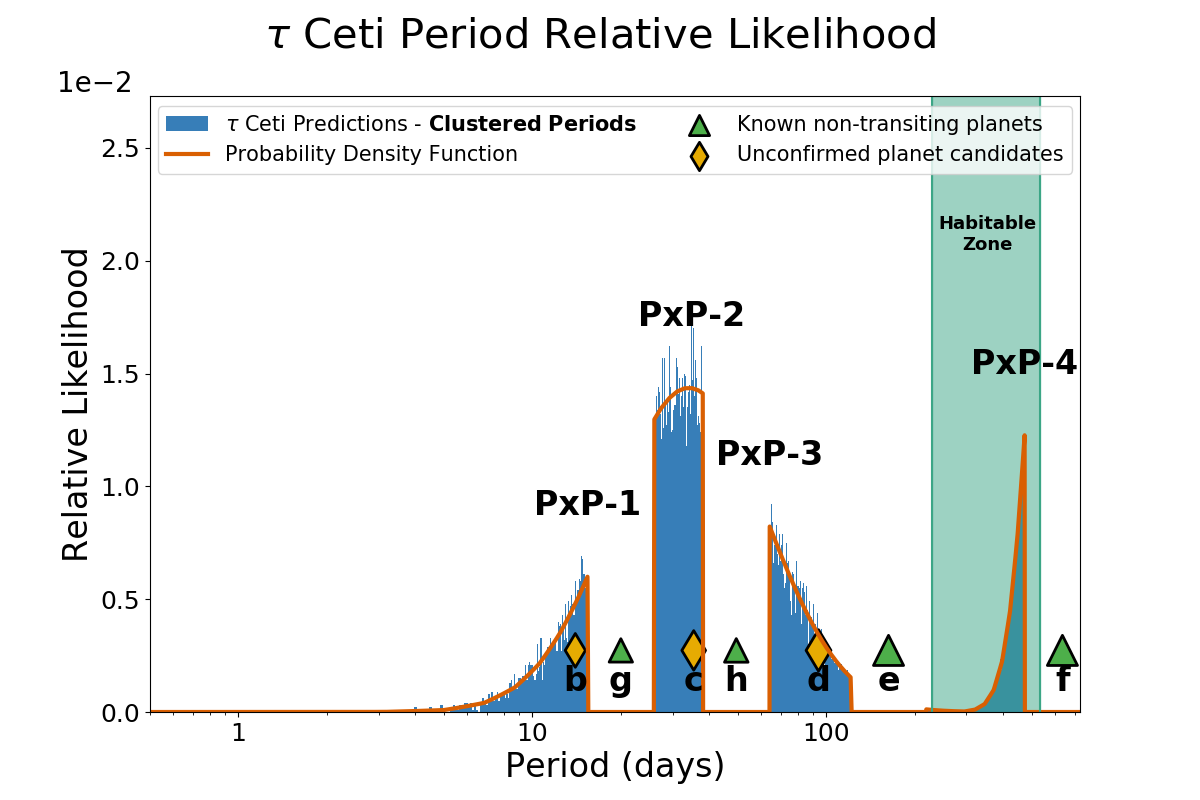}
    \caption{\mname{} predictions for presence of hidden planets around \tauc{} in log period space using the period ratio prescription (top) and clustered periods prescription (bottom).}
    \label{fig:tCP}
\end{figure*}

Under the assumption that the planets are clustered in period space, \mname{} also finds a very similar configuration to that described previously.  Specifically, there are still four relative likelihood maxima for the predicted planets (again named PxP--1 through PxP--4 here) at periods of 14, 34, 67, and 468 days.  These predictions are even closer (in orbital period) for PxP--1 and PxP--2 to planet candidates b and c, but farther away for PxP--3 to planet candidate d.  The cluster assumption means \mname{} finds planet f in a second cluster in period space, away from the main cluster containing the other planets.  Thus, \mname{} predicts the planet as close to the center of that cluster (and therefore as close to planet f) as possible (i.e., dynamically stable).  The relative likelihood for the predicted planets in log-period space, under the clustered periods prescription, is shown in the bottom of Figure~\ref{fig:tCP}.

\subsection{Mass-Radius Relationships}\label{subsec:mr}

On the basis of the \mname{} predictions we will now explore the possible ranges in mass and radius for the planets, planet candidates, and predicted planets in the \tauc{} system.  We will follow three steps: first, we derive the likely mass distributions of the planets.  Second, combining these with planet radius-mass relationships and planet occurrence rates (expressed in planet radii) we derive the likely distributions of planet radii and masses in the system.  Third -- Section~\ref{subsec:nat} -- we will use the derived mass-radius probability distributions to identify the possible natures of the planets, also considering the impact of key assumptions on the results.  In the discussions in this and the following sub-section, we also consider the uncertainties in the measurements as well as the sensitivity of the final results to key assumptions.  The most important assumption is the adopted mass-radius relationship; therefore, we present and contrast results assuming three different mass-radius relationships, which are representative to the state of the art.

For our first step -- deriving the probability distribution functions of the planet masses --  we start our analysis from the $m \sin (i)$ values observed for the confirmed planets and combine these with a constraint on the orbital inclinations to derive the likely true masses of the planets.  The planetary orbital inclinations are not directly measured, but -- as explained in Section~\ref{sec:methods} -- there are very strong astrophysical reasons to assume that the  planetary orbits are overall well-aligned with the debris disk, for which good geometrical constraints exist from spatially-resolved ALMA observations.  With this constraint on the orbital inclinations, we find that all planets and planet candidates in the inner system fall likely fall between $3-7$ $M_\oplus$ in mass.  Table~\ref{tab:planets} shows the derived planet mass ranges and inclinations.

A key assumption in the second step of our analysis is the mass--radius (M--R) relationship adopted for smaller planets, of which multiple somewhat different variants have been proposed in the literature \citep[e.g.,][]{che17, bas17, nin18, ote20}.  Instead of adopting any single M--R relationship, we explore three different relationships, one from \citet{nin18} and two from \citet{ote20}.  Therefore, our exploration of multiple M--R relationships quantifies the impact of this choice on the results.  As one of our three relationships to explore, we chose the non-parametric relationship from \citet{nin18} as it provides a probabilistic measure of the mass and radius.  This relationship conditions its predictions on the known mass/radius data we have gathered from exoplanet systems, where super-Earths/sub-Neptunes have the highest occurrence and therefore would be the most predictive.  However, this conditionality also means there is not a 1:1 relationship between radius and mass, and it also greatly increases the computation time required for deriving mass distributions.  Furthermore, as our second and third M-R relationships to explore, we adopted the two power-law relationships from \citet{ote20} as these match data up to 120 $M_\oplus$ very well.  Based on these, we explore the difference between sub-Neptunes and super-Earths in a region of mass-radius space where these two planet populations overlap.  This combined approach, therefore, also provides an assessment of the possible range of natures for these worlds.

\begin{figure*}[ht]
    \centering
    \includegraphics[width=1.05\columnwidth]{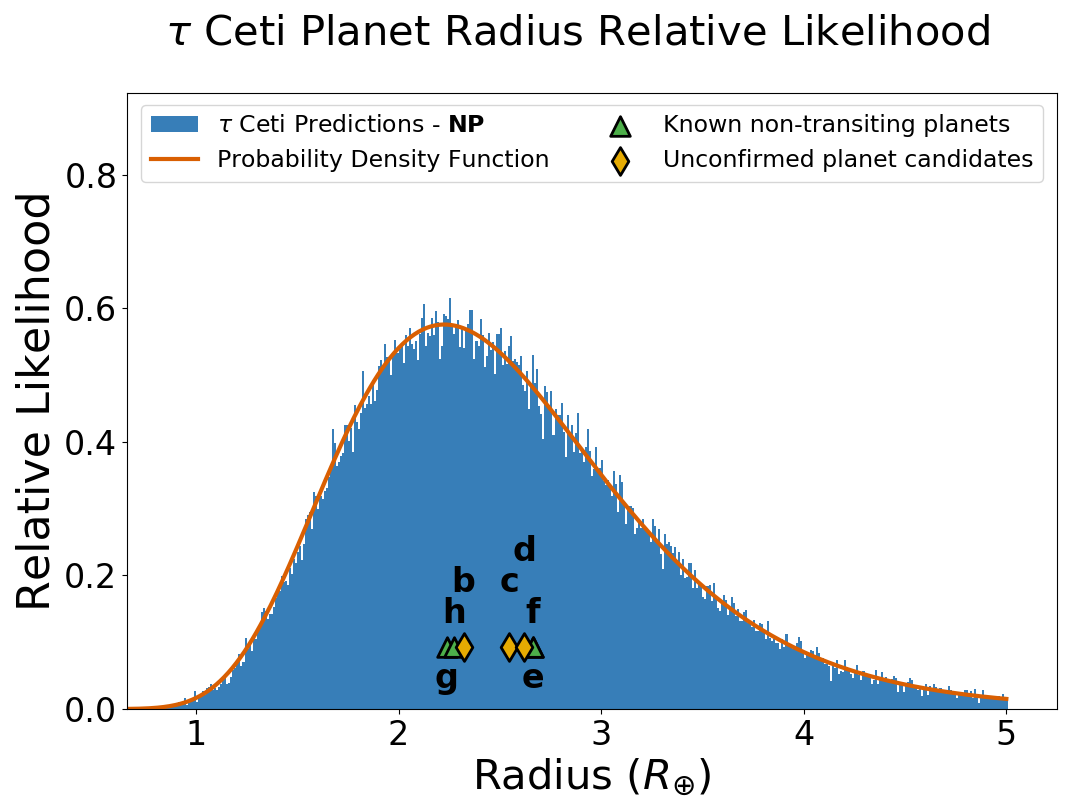}
    \includegraphics[width=1.05\columnwidth]{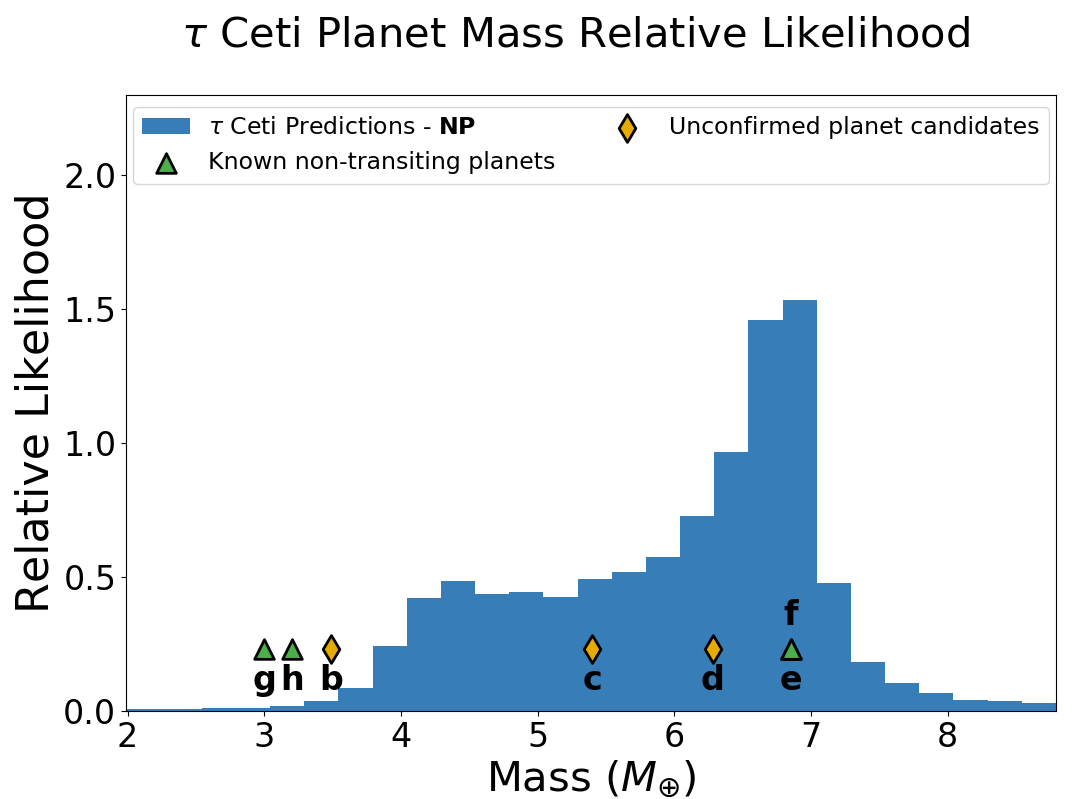}\\
    \includegraphics[width=1.05\columnwidth]{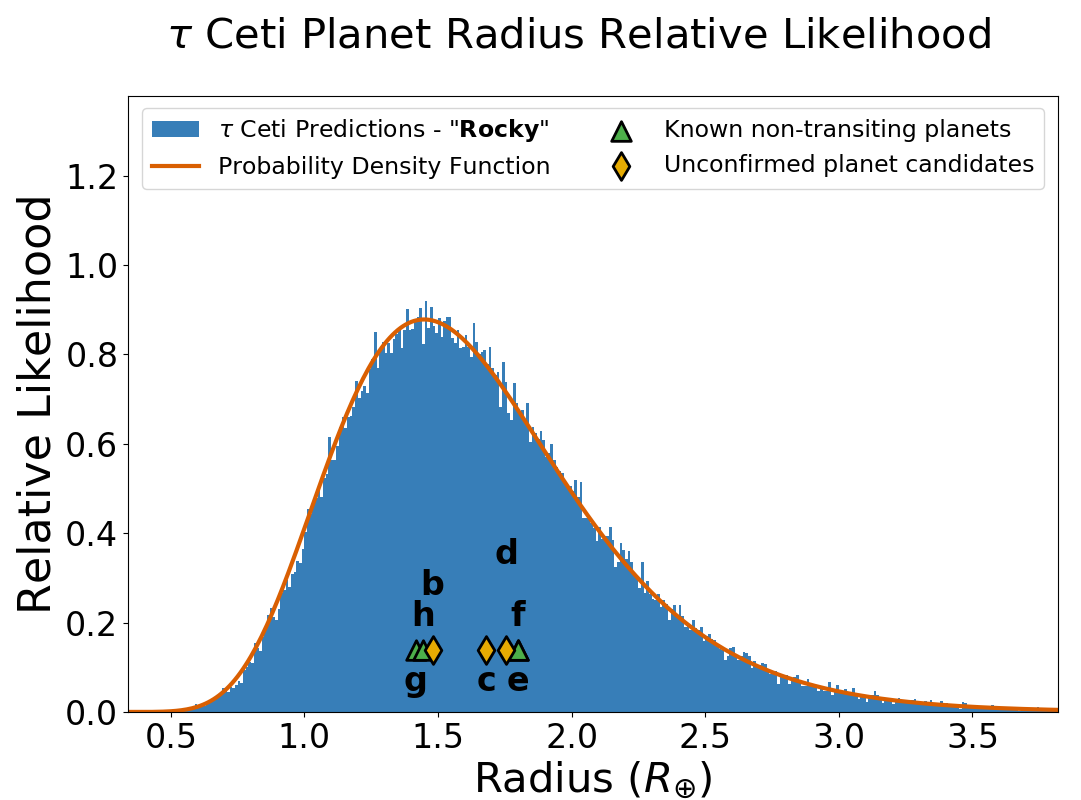}
    \includegraphics[width=1.05\columnwidth]{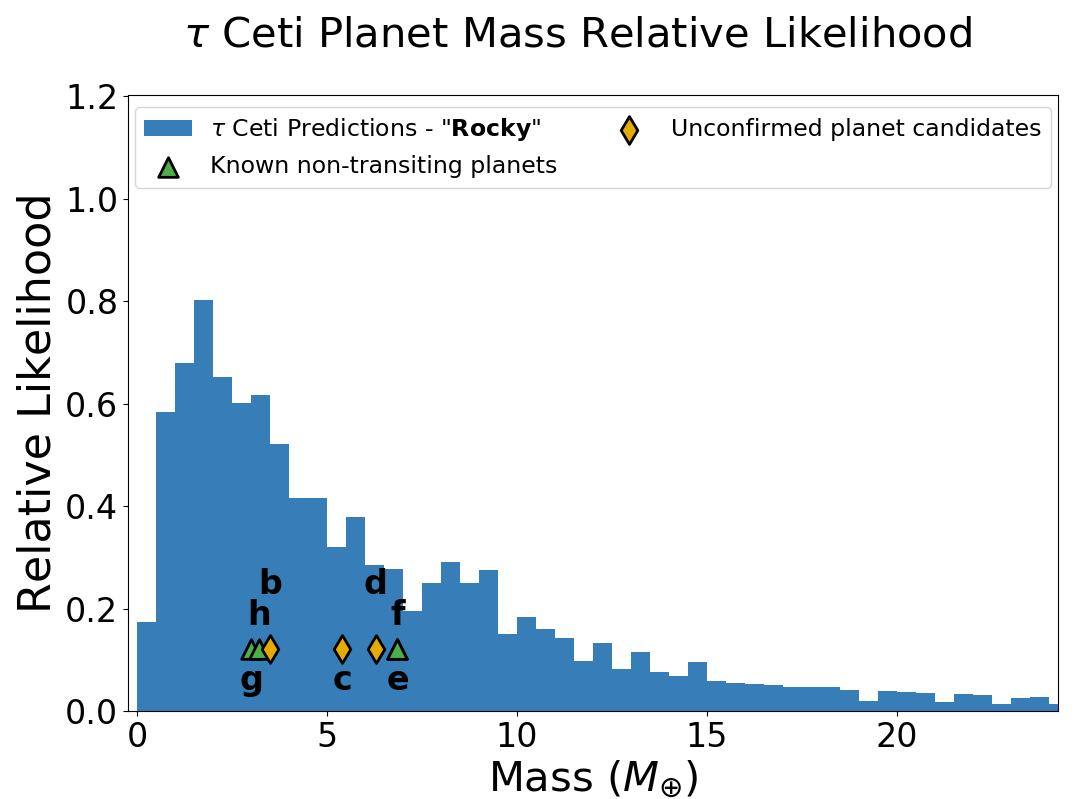}\\
    \includegraphics[width=1.05\columnwidth]{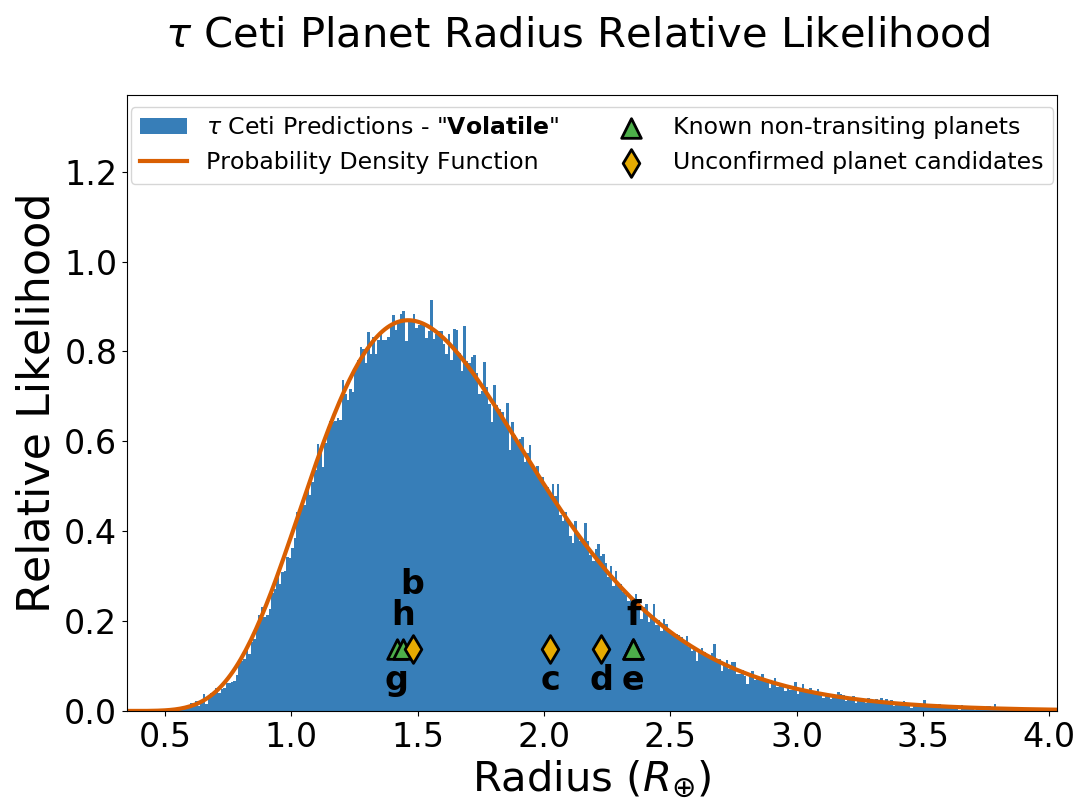}
    \includegraphics[width=1.05\columnwidth]{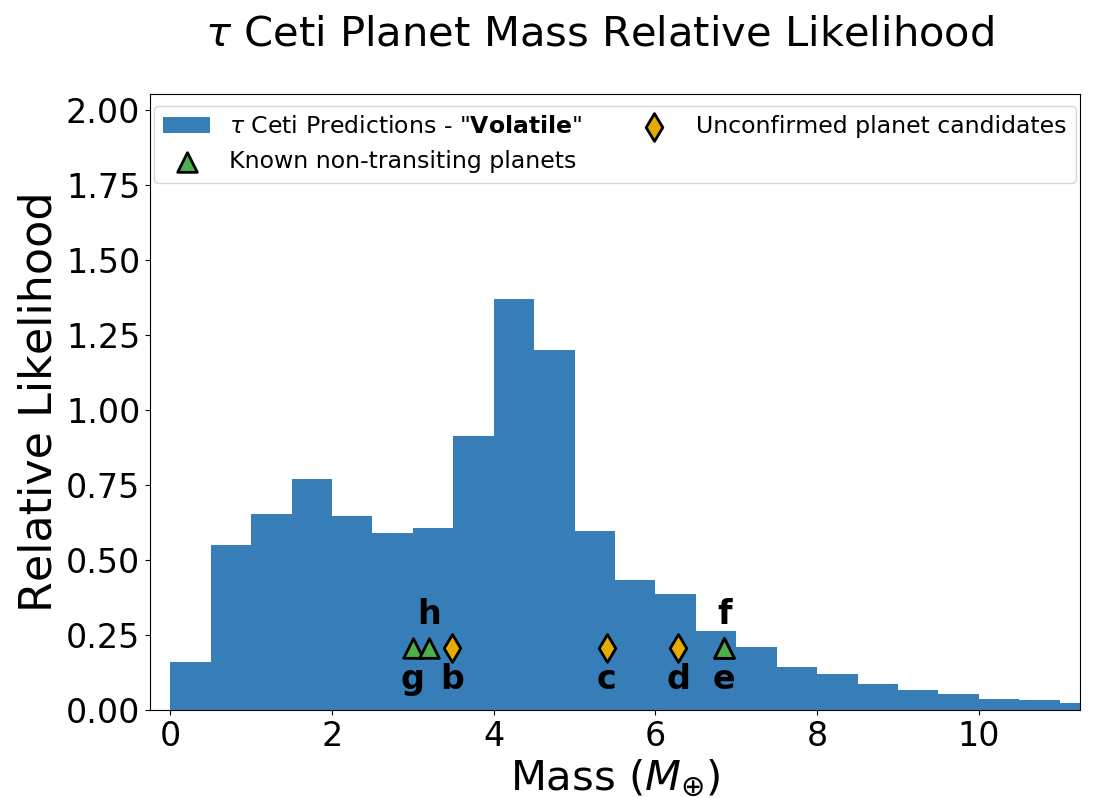}
    \caption{(Left): Planet radius histograms taken from the clustered radius model for three different M--R relationships: \citet{nin18} non-parametric (NP; top), \citet{ote20} power-law ``rocky" (middle), and \citet{ote20} power-law ``volatile" (bottom).  (Right): Planet mass histograms calculated from the radius histograms for the same three M--R relationships.}
    \label{fig:tCRM}
\end{figure*}

We applied each M--R relationship to the known planet masses to find their radii, before -- following  \citet{die20} -- we fit them to the Lognormal distribution for planet radius to find the best-fit distribution.  We find that the known planet radii are close enough to each other to be statistically considered a single cluster in planet radius by \mname{}.  The predicted planet radius distribution found from this best-fit of the four planets, for each of the three M--R relationships adopted, is shown in the left side of Figure~\ref{fig:tCRM}.  We then reversed the process and applied each M--R relationship on the radius distribution to get a distribution in planet masses, shown on the right side of Figure~\ref{fig:tCRM}.

\subsection{The Nature of Planets in the \texorpdfstring{$\tau$\,}{tau }Ceti System}\label{subsec:nat}

In this section we will build on the planet mass--radius probability distributions derived in Section~\ref{subsec:mr} to explore the possible natures of the planets, i.e., by which broad categories they are most likely to fit (e.g., Earth/super-Earth/sub-Neptune/ice giants/evaporated cores).  For our results on the natures of the planets, we simply refer to planet candidates b, c, and d also as planets, as the derived natures would only be meaningful if the planets exist at all.  In the following, we split the planets into two groups based on our mass predictions.

\textbf{Planets b, g, and h}

When using the non-parametric M--R relationship from \citet{nin18}, these three planets would have likely been sub-Neptunes, with predicted radii above $2 R_\oplus$ and a density consistent with a gaseous-volatile envelope.  However, as these planets have expected masses than 5 $M_\oplus$, when using both realizations of the power-law relationships from \citet{ote20}, these planets would be rocky planets, with radii less than 1.5 $R_\oplus$.  Therefore, we find that these worlds are either sub-Neptunes or rocky worlds, depending on which mass-radius relationships better capture reality.

\textbf{Planets c--f}

As with the previous planets, we find that the non-parametric M--R relationship from \citet{nin18} describes these four planets as sub-Neptunes.  The planets have expected masses between $5.4-7$ $M\oplus$, and therefore fall in the overlap region for the \citet{ote20} power-law relationships.  If predominantly ``rocky", these planets would have radii between $1.7-1.8$ $R_\oplus$ and would have approximately terrestrial density. These radii are close in size to the ``Fulton gap'', a relative paucity of planets \citep{ful17}, which may also separate primarily rocky worlds from those with significant gaseous envelopes.  However, the data are also consistent with a different nature: under the ``volatile" M--R relationship from \citet{ote20}, these planets would have planet radii between $2-2.4$ $R_\oplus$, and thus densities and nature comparable to the ice giants in our Solar System.

When comparing the irradiation of the planets in the \tauc{} system to photo-evaporation models \citep[e.g.,][]{owe17, car18}, we find that the three innermost planets b, g, and c would likely have suffered from a significant degree of atmospheric loss due to photo-evaporation.  Therefore, if these planets had any gaseous envelopes, it is likely these atmospheres would have evaporated and left behind a rocky core \citep{pas19}.  However, for planet h and outwards, the radiative flux received from \tauc{} would not be strong enough to strip the envelope from the planet.  Thus it is likely that if these planets had volatile-rich atmospheres, they would still retain them to this day.

\section{Discussion} \label{sec:discussion}

\subsection{Support for previously-identified candidates b, c, and d}

Based on RV-measurements, \citetalias{tuo13} reported five planets in the \tauc{} system (labeled \tauc{} b--f).  However, follow-up analysis by \citetalias{fen17} found four planets in the system.  They discovered new planets at 20 and 49 days (labeled \tauc{} g and h) and confirmed the presence of planets e and f at 163 and 636 days, but were not able to confirm the presence of planet candidates b--d at 14, 35, and 94 days.  They found that the 14-day period signal becomes much less significant in later datasets when the 20-day signal is subtracted, and is, thus, attributed to stellar activity.  Furthermore, the 35.4-day period signal was found to be much weaker than previously thought and close to the rotation period of the star.  Finally, the 94-day period signal was noticeable but found in only a few of the datasets.

To assess the likelihood of planet candidates b--d in the system, we show that if planets were to be added within the orbit of planet e (P $\lesssim$162 days), the periods for \tauc{} b--d are near the local maxima of the relative likelihood in period space.  This is true for both exoplanet period distributions; the clustered periods prescription is closer to the known signals for b and c, while the period ratio prescription is closer for d.  Therefore, the predictions for planet candidates PxP--1, PxP--2, and PxP--3 provide contextual, statistical evidence that support the existence of planet candidates \tauc{} b--d, as does the dynamical stability analysis of the \tauc{} system.  When only considering planets g, h, e, and f in the system, the average separation between each planet in units of mutual Hill radii is $\sim$31.4 (assuming all planets orbit near the disk inclination).  The peak in separation seen in population statistics is $\sim$20 mutual Hill radii, with larger numbers (i.e., close to 40) indicating the presence of missing planets \citep[e.g.,][]{gil20}.  Adding in \tauc{} b--d and PxP--4 with the period ratio description pushes the average separation down to $\sim$15.6 Hill radii.

\subsection{PxP--4 characteristics -- A Habitable Zone Super-Earth?}

Both flavors of the orbital period distributions explored in our analysis results in the robust prediction that a planet in the habitable zone of \tauc{} is very likely.  This prediction is, as discussed below, consistent with available data.
\citetalias{tuo13} identifies a possible $\sim$315 day signal and performed extensive Keplerian modeling of the system both with and without that planet.  They concluded that it was more likely an alias of the 168 day signal.  \citetalias{fen17} also found a relatively strong signal at $\sim$318 days and concurred that it was an alias of their 162-day signal.  However, due to the gap of factor $\sim$4 in period space between planets e and f, \mname{} (and dynamically packing, in general) predicts another planet candidate, PxP--4, in the period range between those two known planets.

With the four-planet architecture found by \citetalias{fen17}, under the clustered periods assumption, our analysis finds that planet f is separated from the other planets in period space and is in its own cluster.  Therefore, we predict another planet as close to the center of that cluster (i.e., the period of planet f) as dynamically possible, near orbital period 468 days.  In contrast, under the period ratio assumption, our analysis places the new planet with symmetric period ratios between itself and planets e and f, at a period of 322 days.  Given the mass of \tauc{} as $0.783$ $M_\odot$, an orbital period of 322 days corresponds to an orbital semi-major axis of 0.848~au, and an orbital period of 468 days corresponds to orbital semi-major axis of 1.09~au.  The conservative habitable zone for \tauc{} (effective temperature $T_{\mathrm eff} = 5,344$ K), as defined by \citet{kop13} using their ``Moist Greenhouse" and ``Maximum Greenhouse" limits, lies between stellocentric distances of $0.703-1.26$~au.  Thus, PxP--4 would comfortably lie inside the habitable zone of \tauc{}.

As before, we explore the potential nature of PxP--4 by contrasting it to the exoplanet size--density trends.  As explained below, we find that it may be a super-Earth, but could also be a volatile-rich sub-Neptune.  We calculated the mass of the PxP--4 planet candidate by translating its predicted planet radius distribution into a mass distribution via our M--R relationship.  This provided the predicted mass of 3.33 $M_\oplus$ for the power-law relationships from \citet{ote20} and 6.30 $M_\oplus$ for the non-parametric M--R relationship from \citet{nin18}.  Planets of the former mass (3.33 $M_\oplus$) would likely always be rocky, as the lower limit for the volatile-rich population from \citet{ote20} where anything below is considered rocky is 5 $M_\oplus$.  Under the ``rocky" assumption, 1 $M_\oplus$ lies at the 7th percentile of the planet mass distribution for the \tauc{} system.  Planets with the latter mass ($M_p \approx 6-7$ $M_\oplus$) would have planet radii $\sim$1.8 $R_\oplus$ if they were rocky, but could also likely contain a gaseous envelope and have radii of $2.1-2.4$ $R_\oplus$.

\subsection{Observational signatures of PxP--4}

Our prediction of a habitable zone planet, PxP--4, in the \tauc{} system raises the question of whether this planet could be detected.  In this section we explore the observational signatures expected from this planet and discuss what measurements may clarify whether it is a habitable zone super-Earth or a -- presumably uninhabitable -- sub-Neptune.  Specifically, we will discuss the possibility of detecting PxP--4 via transits, radial velocity, and direct imaging.  Our analysis assumes that the inclination of PxP--4 is aligned with the disk and other planets in the system to within a few degrees.  This assumption would make it very unlikely that PxP--4 is detectable via transit observations, leaving RV and direct imaging as the viable options for the next decades of observations.  

To estimate the RV semi-amplitude, we assume that PxP--4's orbital eccentricity is low, as would be expected in a high-multiplicity system \citep[e.g.,][]{lim15, van19}.  Given its value for $m \sin i$  of $3.3-6.3$ $M_\oplus$, the RV semi-amplitude of PxP--4 would be $\sim$0.2--0.4 m/s, which is similar in magnitude to the signals detected from \tauc{} g and h.  This value is at or just above the noise limit for \tauc{} on the High Accuracy Radial Velocity Planet Searcher \citep[HARPS;][]{may03} spectrograph, as characterized by \citetalias{fen17}.  The ESPRESSO spectrograph \citep{pep10} would likely be able to reach the required precision to detect PxP--4, as it is able to reach $\sim$0.25 m/s for Proxima Centauri \citep{sua20}, a fainter and more active star.

We will now explore \tauc{}'s PxP--4 as a potential target for direct imaging.  \tauc{}'s relative proximity to the Solar System makes this system one of the premier targets for next-generation direct imaging systems, both from space and from ground.  A direct imaging detection of a potentially habitable PxP--4 would be revolutionary, as it could be exploited for a broad variety of follow-up efforts to characterize the planet in depth and to address a multitude of new science questions \citep[for a community report, see][]{apa17}.  At a distance of only 3.65~pc, the orbital semi-major axis of 0.848~au of PxP-4 corresponds to an angular separation of 250~mas, which is on par to recently imaged super-Jupiter exoplanets \citep[e.g.,][]{lag10,mac15}.  Unlike the hot young super-Jupiters imaged with extreme adaptive systems, however, PxP--4 is neither young, hot, or massive, and thus poses an orders-of-magnitude greater contrast challenge than anything imaged as of now.

While imaging PxP--4 with current state-of-the-art facilities is not possible, the feat may be within grasp of future facilities.  Thermal emission from PxP--4 should peak close to $\lambda_{\mathrm{th}}=10\,\mu$m due to the equilibrium temperature of this habitable zone planet.  Current efforts for detecting a habitable zone Earth-sized planet around $\alpha$~Centauri through a 100~h-long integration with the Very Large Telescope's VISIR instrument show it is possible \citep[e.g.,][]{kas19}.  Although \tauc{} is at a 2.4$\times$ greater distance than $\alpha$~Centauri, PxP--4 may be a factor of a few larger than Earth, somewhat countering the greater distance.  Thus, while the intensity of PxP--4 may be detectable with the sensitivity of present-day, ultra-deep thermal infrared images \citep[see also][]{wag20}, its relatively small projected separation (250~mas) falls within the contrast-limited region of such images, where sensitivity is greatly reduced (Wagner et al., in prep.).  The next-generation large ground-based telescopes, such as the Giant Magellan Telescope\footnote{GMT Science Book, \url{https://www.gmto.org/sciencebook2018}}, the Thirty Meter Telescope, or the European Extremely Large Telescope, should deliver increased thermal infrared sensitivity and a factor of several smaller inner working angles \citep[e.g.,][]{qua15, maz19, wan19}, making direct detection of \tauc{} PxP--4 a promising possibility.

PxP--4 may also be detected in the future via reflected (scattered) light imaging.  The brightness of the planet in scattered light depends on the physical separation from the host star, the Bond albedo, the scattering phase angle, and the illumination phase, which complicate the detection, identification, and interpretation of directly imaged habitable planets \citep[e.g.,][]{bix20}.  It is likely that PxP--4 would be $\sim$10$^{-9}$ fainter than \tauc{}, requiring space-based, ultra high contrast imaging or interferometry to detect.  Mission concepts that would be capable of detecting PxP--4 have been proposed (e.g., LUVOIR: \citealt[][]{luv19}, HabEx: \citealt{gau20}, and LIFE: \cite{qua19}), and may become operational within the next two decades.

Thus, we conclude that PxP--4 is already likely detectable indirectly, via the RV-modulations it imprints on its host star, and it will be directly detectable by the end of the current decade in thermal emission (with the extremely large ground-based telescopes), as well as an ideal target for space-based high-contrast imaging and spectroscopy in about two decades.

\subsection{Habitable Planet in \texorpdfstring{$\tau$\,}{tau }Ceti System?}

\tauc{} has long been a candidate to search for nearby life, and we find that the probability of having a rocky planet in the habitable zone is relatively high.  The probability of additional planets to be sub-Neptunes with a gaseous/volatile envelope vs. rocky super-Earths ranges from roughly equal to leaning towards rocky planets; across the M--R relationships $50-75$\% of injected planets had a planet mass below 5 $M_\oplus$.  However, the probability of finding specifically an Earth-like planet (0.5 $M_\oplus < M_p <$ 1.5 $M_\oplus$) is low.  The non-parametric M--R relationship from \citet{nin18} had no significant probability of finding a planet with a mass between $0.5-1.5$ $M_\oplus$, whereas the rocky and volatile populations of the power-law relationships from \citet{ote20} had only $\sim$12\% and $\sim$3\%, respectively, in that mass range.

A super-Earth planet in the habitable zone around \tauc{} would have a RV semi-amplitude of 0.4 m/s, which is above the noise floor of current spectrographs like HARPS and ESPRESSO.  However, even across a decade of observations, this signal would still be difficult to find, especially due to the alias caused by Earth's own orbital period.  With even better extreme-precision radial velocity (EPRV) measurements pushing down towards 1 cm/s precision, the ability to observe a signal from rocky planets in the habitable zone around \tauc{} will become much easier, and EPRV data would even be able to find an Earth-like planet (RV semi-amplitude $\lesssim 0.1$ m/s) around \tauc{}.

\begin{figure*}
    \centering
    \includegraphics[width=2.12\columnwidth]{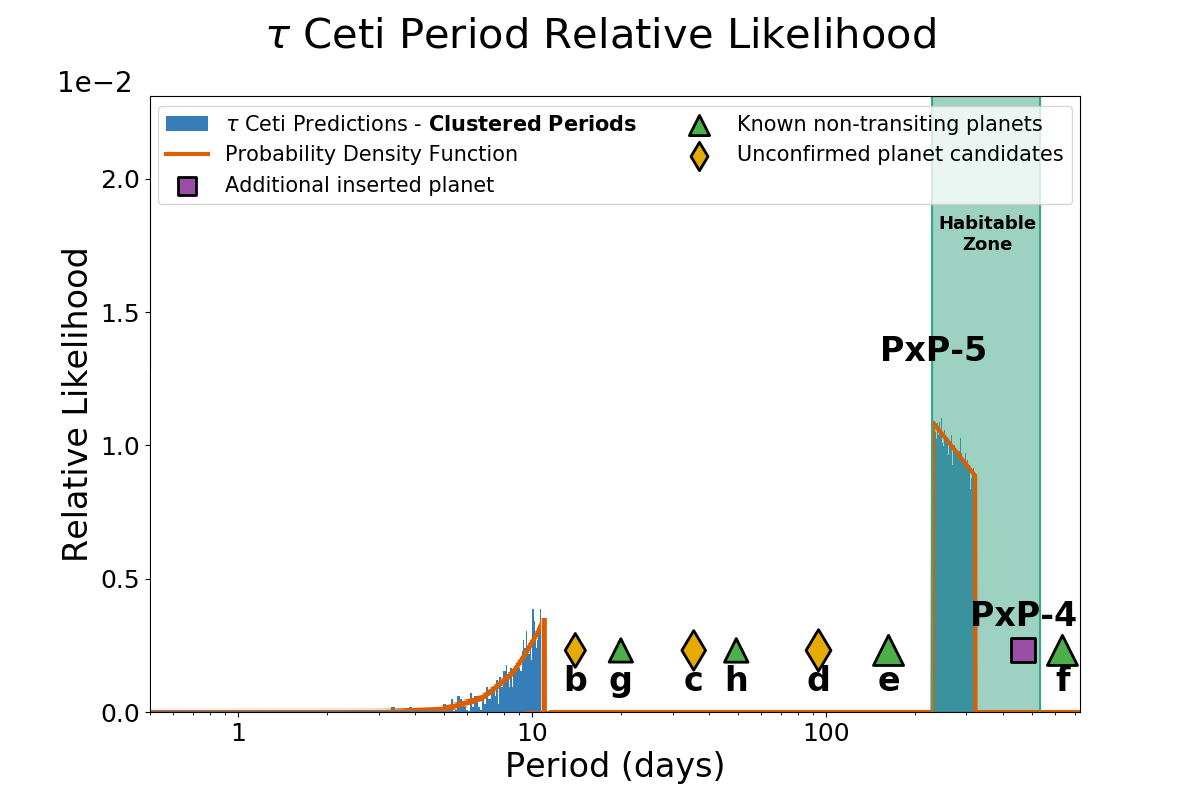}
    \caption{The clustered periods prescription predictions after adding in all 4 predicted exoplanets.  The gap between e and the additional inserted planet PxP--4 is large enough for another planet to fit in between, with a period of $\sim$270 days, at the inner edge of the habitable zone.}
    \label{fig:syssim_add_all}
\end{figure*}

The period ratio prescription has low probability of adding in an additional planet after inserting PxP--1-3 at the values for the known planet candidates b-d and adding PxP--4 at its relative likelihood maxima.  These four predicted planets dynamically pack the system with relative period ratios for each pair of planets close to the Kepler statistical mean.  The clustered periods prescription, after adding in PxP--1-3 at the values for planet candidates b-d and PxP--4 at its relative likelihood maxima, also allows for an additional planet candidate to be placed at $\sim$270 days, corresponding to an orbital semi-major axis of 0.754~au.  This hypothetical planet, too, would be within the habitable zone assuming an inner limit of 0.703~au.  This additional prediction from the clustered periods prescription is favorable as it fits the system parameters well.  The integrated likelihood of the second injection is higher than the first injection, and the overall average planet separation in units of mutual Hill radii for the nine planet system prediction via the clustered periods prescription is 15.5.  This is very similar to the eight-planet system prediction via the period ratio prescription of 15.6.  The new injection test, after adding in the eight current candidates, is shown in Figure~\ref{fig:syssim_add_all}.

\section{Summary} \label{sec:summary}

We applied the \mname{} algorithm to provide an integrative analysis of the \tauc{} system, the second-closest Sun-like star to the Solar System.  The key findings of our study are as follows:

(1) Using both orbital period distribution models, our analysis suggests that the three planet candidates (b, c, d) reported by \citetalias{tuo13} at periods of $\sim$14.0, 35.4, and 94.1 days are likely to be real.  These planets would dynamically pack the system and follow the \textit{Kepler} population statistic for the first planet being found at an orbital period mode of 12 days \citep{mul18}.

(2) Our analysis predicts an additional, fourth planet candidate (PxP--4) at P = 322 days (when using the period ratio prescription) or at P = 468 days (using the clustered periods prescription).  The period ratio prediction would fully pack the system out to known planet f, and matches a possible signal in the RV data seen by both \citetalias{tuo13} and \citetalias{fen17} at 315-320 days.  The clustered periods prediction also allows for another planet candidate between planet e and PxP--4.

(3) Given the assumption that the orbital plane of the planets matches the visible debris disk at 35$^\circ$, the planets and planet candidates are all super-Earths or sub-Neptunes ($3-7 M_\oplus$).  Similarly, we predict candidate PxP--4 to have a planet mass between $3.3-6.3 M_\oplus$.  The probability that this candidate is rocky ($M_p < 5 M_\oplus$) is more than 50\%, but the likelihood of it being Earth-like ($0.5 M_\oplus < M_p < 1.5 M_\oplus$) is $\lesssim 10\%$.

(4) With \tauc{} being Sun-like but only half as luminous, if PxP--4 orbits \tauc{} with a 320-470 day period (or an orbital semi-major axis of $\sim0.85-1.09$~au), it would receive roughly 35-60\% of the light that Earth does.  Thus, PxP--4 would straddle the center of \tauc{}'s habitable zone \citep{kop13}.

(5) The low end of the predicted mass range for PxP--4 is near the masses of \tauc{} g--h and PxP--2/planet b ($3-3.5 M_\oplus$).  The high end of the predicted mass range for PxP--4 is near the masses of \tauc{} e--f ($\sim 7 M_\oplus$) and similar to the mass of PxP--3 (a close match to the unconfirmed planet candidate d).  Planets g and h, and planet candidate b are likely rocky, while planets e and f, along with planet candidates c and d, each have a roughly equal likelihood of being rocky or containing a significant gaseous envelope.

(6) The predicted presence of PxP--4 in the habitable zone should be soon testable.  While the RV semi-amplitude for PxP--4 of $0.2-0.4$ m/s is at the precision limit for \tauc{} of the current data, PxP--4 is likely to be within reach of the newest and near-future EPRV instruments.  With the new generation of extremely large telescopes (GMT, TMT, and E-ELT), high-contrast thermal infrared imaging should enable direct detection and study of this world.  Futhermore, this potentially habitable planet could be studied in great details with future space-based telescopes, such as the mission concepts LUVOIR, HabEx, or LIFE.

(7) If PxP--4 is close to the widest predicted orbits (i.e., has a period close to $\sim$470 days), we find that an additional planet may reside in the habitable zone.  This second habitable zone planet would then have a period of $\sim$270 days.

Our study demonstrates an approach to exploring the inner planetary systems of nearby stars.  We combine uncertain and incomplete but specific information on planetary systems with robust statistical understanding of exoplanet population demographics and first principles-based constraints on planetary dynamics. The current \mname{} implementation is computationally inexpensive and can be applied to a large sample of systems.  The analysis done in this work and as carried out by \citet{die20} on the TESS sample of multi-planet systems takes a couple of minutes per system on a 16-core desktop computer, for every setup with the exception of the non-parametric model from \citet{nin18} as stated in Section~\ref{subsec:nat}. For the \tauc{} system, the integrated analysis supports the veracity of three planet candidates reported in the literature and predicts the presence of a habitable zone, possibly rocky planet (PxP--4).  Soon, improved radial velocity coverage should be able to directly test the predictions made in this study and possibly confirm the presence of PxP--4 in the habitable zone.  This measurement will represent a great leap into clarifying the potential of the second-closest Sun-like star and closest single Sun-like star -- an obvious target for biosignature searches -- for hosting a habitable world.

\acknowledgments

We acknowledge support from the Earths in Other Solar Systems Project (EOS), grant no. 3013511 sponsored by NASA.  The results reported herein benefitted from collaborations and/or information exchange within NASA's Nexus for Exoplanet System Science (NExSS) research coordination network sponsored by NASA's Science Mission Directorate.  We acknowledge use of the software packages NumPy \citep{har20}, SciPy \citep{vir20}, Matplotlib \citep{hun07}, and MRExo \citep{kan19}.  This paper includes data collected by the Kepler mission.  Funding for the Kepler mission is provided by the NASA Science Mission Directorate.  The citations in this paper have made use of NASA’s Astrophysics Data System Bibliographic Services.

\bibliography{main}

\end{document}